# Zero-Field In-Plane Critical Current Density of $YBa_2Cu_3O_{7-\delta}$ Thin Films

***M. Rakibul Hasan Sarkar*[1] *and S. H. Naqib*[2*]**

[1]*Department of Science and Humanities, Military Institute of Science and Technology, Mirpur Cantonment, Dhaka-1216, Bangladesh*
[2]*Department of Physics, University of Rajshahi, Rajshahi-6205, Bangladesh*

## Abstract

Temperature dependence of the zero-field critical current density, $J_{c0}(T)$, in the $CuO_2$ planes has been investigated for *c*-axis oriented $YBa_2Cu_3O_{7-\delta}$ (Y123) thin films with different hole content, *p*. *p*-values were varied by changing the oxygen deficiency, $\delta$. $J_{c0}(T)$ were calculated from the magnetic hysteresis (*M-H*) loops obtained at different temperatures. From the analysis of the $J_{c0}(T)$ data for different sample compositions, we have extracted valuable information regarding the nature of the underlying magnetic flux pinning mechanisms in Y123 superconductors. It is found that the oxygen defects play only a secondary role in pinning of the vortices, the superfluid density, on the other hand plays a significant role. The temperature exponent, *n*, governing the $J_{c0}(T)$ behavior for a given sample composition showed a systematic behavior as hole contents are changed.

## 1. Introduction

These days the double $CuO_2$ layer $YBa_2Cu_3O_{7-\delta}$ is considered as one of the most promising systems for high critical current density applications among all the cuprate high-$T_c$ superconductors [1 – 3]. The relatively higher superfluid density and robust interlayer coupling also make the irreversibility magnetic fields higher in Y123 and related R123 (R ≡ Rare earth) compounds [4, 5]. The superconducting critical current density is limited by both intrinsic and extrinsic effects present in the compound of interest. For example, $J_c$ is limited by both the intrinsic depairing effect and also by the depinning mechanisms, which is largely related to the nature of the extrinsic defects present inside the sample. These defects act as magnetic flux pinning centres in type-II superconductors. The depairing effect is due to the breaking of Cooper pairs by the induced supercurrent. The depinning effect, on the other hand, is governed by the interplay between magnetic

---

**Corresponding author*: salehnaqib@yahoo.com

flux dynamics and flux pinning mechanisms [6]. The intrinsically high depairing critical density is significantly reduced due to the presence of extrinsic effects (*e.g.*, weak links and grain mismatch). Understanding the temperature and composition dependences of these effects are of great importance in order to design superconductors capable of supporting large critical current density and irreversibility magnetic fields.

In this study we have analyzed the temperature dependence of the zero magnetic field *ab*-plane critical current density, $J_{c0}(T)$, of high-quality *c*-axis oriented thin films of $YBa_2Cu_3O_{7-\delta}$ compounds with different oxygen deficiencies and therefore, at different levels of hole contents. For a given compound at a fixed temperature the zero-field critical current density is the maximum current density and large scale power applications depend on this parameter.

The oxygen deficiency, $\delta$, determines both the number of extra holes, $p$, in the $CuO_2$ planes and the level of disorder in the $CuO_{1-\delta}$ chains. In high-$T_c$ cuprates, the intrinsic parameters are controlled by the hole content [7], whereas disorder leads to the extrinsic effects. From the analysis of the $J_{c0}(T)$ data we have found that the critical current density is mainly governed by the superfluid density (related to the depairing current) and the chain oxygen defects play only a minor role.

This paper is organized as follows. In section 2 we have presented the experimental details. Section 3 consists of the analysis of the $J_{c0}(T)$ data. Finally, in section 4 we have discussed the implications of our main findings from the analysis. This section also includes the conclusion that can be drawn from the discussions.

## 2. Experimental samples and results

High quality epitaxial *c*-axis oriented thin films of $YBa_2Cu_3O_{7-\delta}$ were fabricated using the method of pulsed *LASER* deposition (PLD). High density (> 90% of ideal X-ray density) sintered $YBa_2Cu_3O_{7-\delta}$ was used as the target material. The films were grown on crystalline $SrTiO_3$. The substrates were highly polished on one side and were 10 mm x 5 mm x 1 mm in size. The crystallographic orientation of the substrate was (100) ($SrTiO_3$ is a cubic system with a lattice constant of 3.905 Å) [8].

The films were fabricated using a *Lambda Physik LPX210* KrF *LASER* with a wavelength of 248 nm. The laser intensity inside the deposition chamber was measured before each PLD run and was adjusted to the desired value using focusing/attenuating systems. A pulse rate of 10 Hz was used during thin film deposition. Prior to the deposition, the substrate heater block was fully out-gassed by heating it to the deposition temperature with a turbo-pump running until the chamber pressure reached ~ $10^{-5}$ mbar. The deposition temperature, $T_{ds}$, was measured by a thermocouple fixed to the heater block. A $T_{ds}$ of 780$^{o}$C was used for the Y123 films. The oxygen partial pressure lied within the range from 0.70 mbar to 1.00 mbar during deposition. The average thickness of the films used in this study is (2700 ± 300) Å. All the films were characterized by X-ray diffraction (XRD), resistivity, room-temperature thermopower (RT-TEP), and atomic force microscopy (AFM). XRD gave information regarding crystalline structure, phase purity, degree of epitaxial growth, and oxygen deficiency (via *c*-axis lattice parameter [9]). From the in-plane resistivity, $\rho_{ab}$, we obtained information regarding disorder content inside the films and the superconducting transition temperature, $T_c$. From RT-TEP, the hole content was extracted [9 – 11]. AFM, on the other hand gave information about the surface quality and size of the crystalline grains. Details of film characterization can be found in ref. [12]. The XRD pattern of the as-grown film is shown in Fig. 1. Fig. 2 shows the $\rho_{ab}(T)$ data for the same sample. Information regarding the oxygen annealing and hole content is presented in Table 1.

Table 1: PLD and oxygen annealing conditions, oxygen deficiency ($\delta$), superconducting transition temperature ($T_c$) and hole content ($p$) of the thin films.

| **Sample** | **Annealing Conditions** | **$\delta$ (±0.03)** | ***p* (±0.004)** | **$T_c$ (±1.0) (K)** |
|---|---|---|---|---|
| $YBa_2Cu_3O_{7-\delta}$ | AP: As prepared. $T_{ds}$ = 780$^0$ C, $PO_2$ = 0.95 mbar. Rapidly cooled *in-situ* down from $T_{ds}$ to 450$^0$C in $O_2$ before rapid cooling to room temperature. | 0.10 | 0.162 | 90.5 |
| | 575$^0$ C in $O_2$ for 2 hrs, furnace cooled to room temperature. | 0.20 | 0.147 | 89.8 |
| | 650$^0$ C in $O_2$ for 2 hrs, furnace cooled to room temperature. | 0.31 | 0.104 | 64.0 |

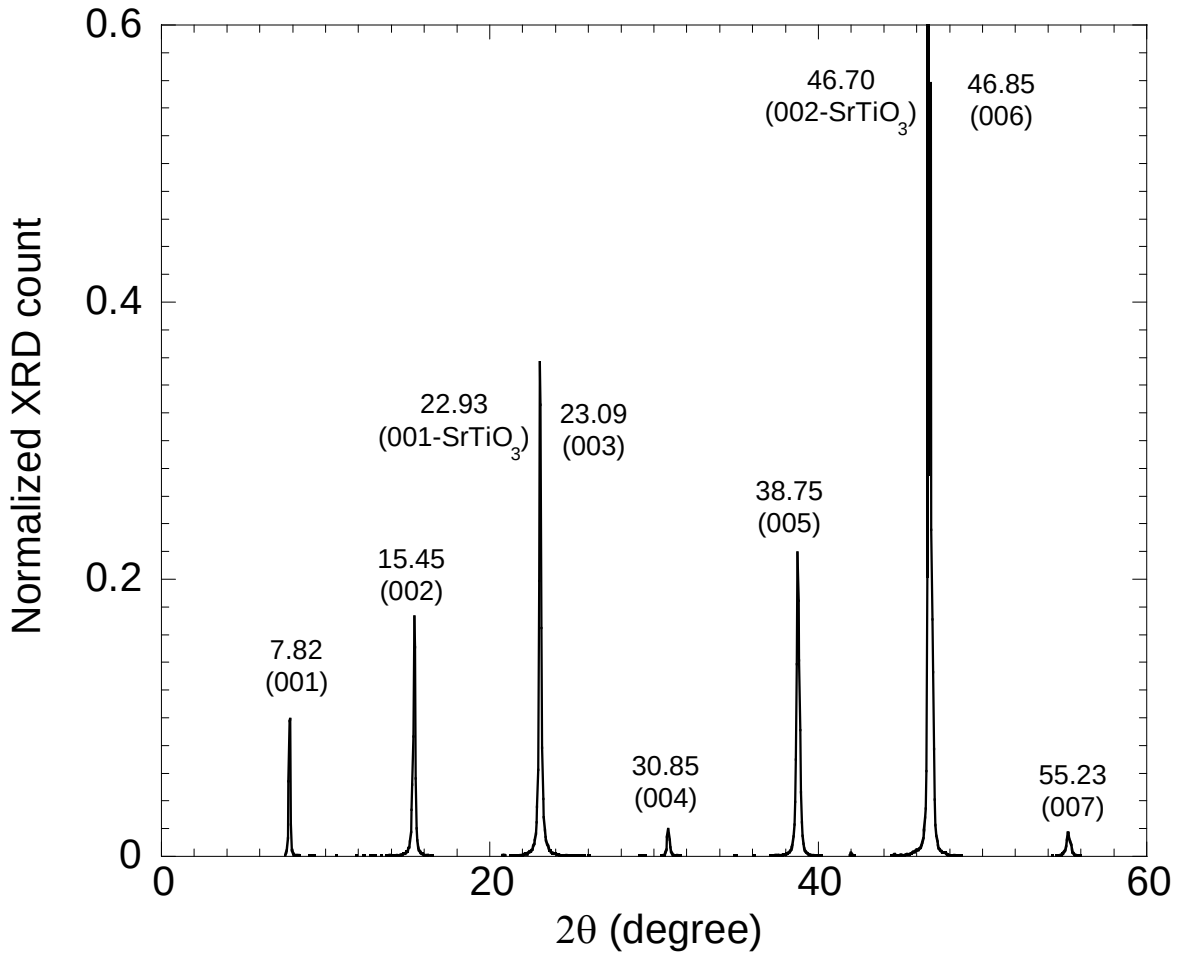

Fig. 1. X-ray diffraction pattern for the as-prepared Y123 thin film. The reflection planes are given inside the brackets. The angles are also marked. For clarity of the less intense peaks, only part of the (006) peak has been shown.

The expected XRD pattern for a *c*-axis oriented film would be a series of (00*l*) peaks, without any missing reflections due to imperfect structure of the unit cell. Very clear and sharp sets of (00*l*) peaks were obtained by the XRD, without any trace of impurity phase, as shown in Fig. 1. $\rho_{ab}(T)$ data for the experimental samples (*e.g.*, Fig. 2 shows the plot for the as-prepared thin film) are characterized by small residual resistivity and sharp superconducting transition, indicating that the bulk of the films are largely defect free and homogeneous.

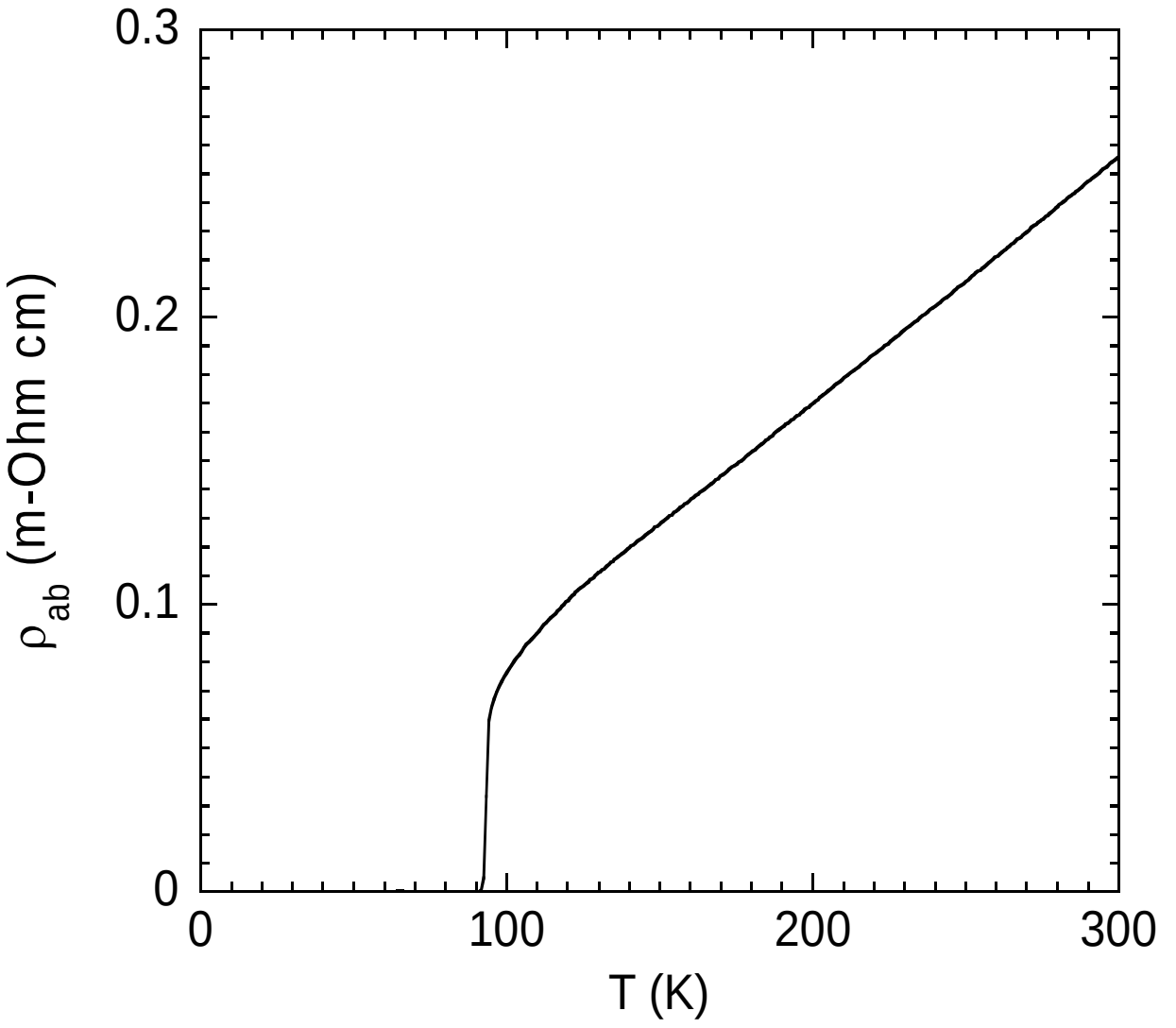

Fig. 2. In-plane resistivity of the as-prepared, optimally doped, Y123 thin film after *in-situ* oxygen annealing (details in Table 1).

$J_c$ was calculated from the magnetic hysteresis loops at different temperatures. The field dependent magnetization (*M-H* loop) was measured using a vibrating sample magnetometer (VSM). An *EG&G PAR* VSM was used. The magnetic field ramp rate was 500 Oe/min during the hysteresis measurement with the applied magnetic field always perpendicular to the surface of the *c*-axis oriented thin films, *i.e*, $H \| c$ in all cases. In this configuration the supercurrent flows in the *ab*-plane. All the *M-H* loops were obtained at various fixed temperatures below $T_c$. Temperature was stabilized within ±1 K during measurements. The hysteresis (*M-H*) loops for the samples under study are shown in Figs. 3. The Bean Critical State model [13] provides the link between $J_c$ and hysteretic magnetization. But the Bean model is appropriate only for certain type of sample geometry (*e.g.*, infinite sheet with parallel magnetic field). For finite geometry with the magnetic field perpendicular to the plane of the sample (as is the case for our thin films), one has to use a modified Critical State model. For rectangular samples Brandt and Indenbom found the following expression for $J_c$ [14]

$$J_c(H) = \frac{\Delta M(H)}{2a^2 lt} \quad (1)$$

Where, $\Delta M(H)$ is the difference between upper and lower branches of the *M*-*H* loops at a given field and temperature. $2a$, $l$ and $t$ are the width, length and thickness of the thin films, respectively. Fig. 4 shows the $J_{c0}(T)$ plots with different values of $p$ ($\delta$).

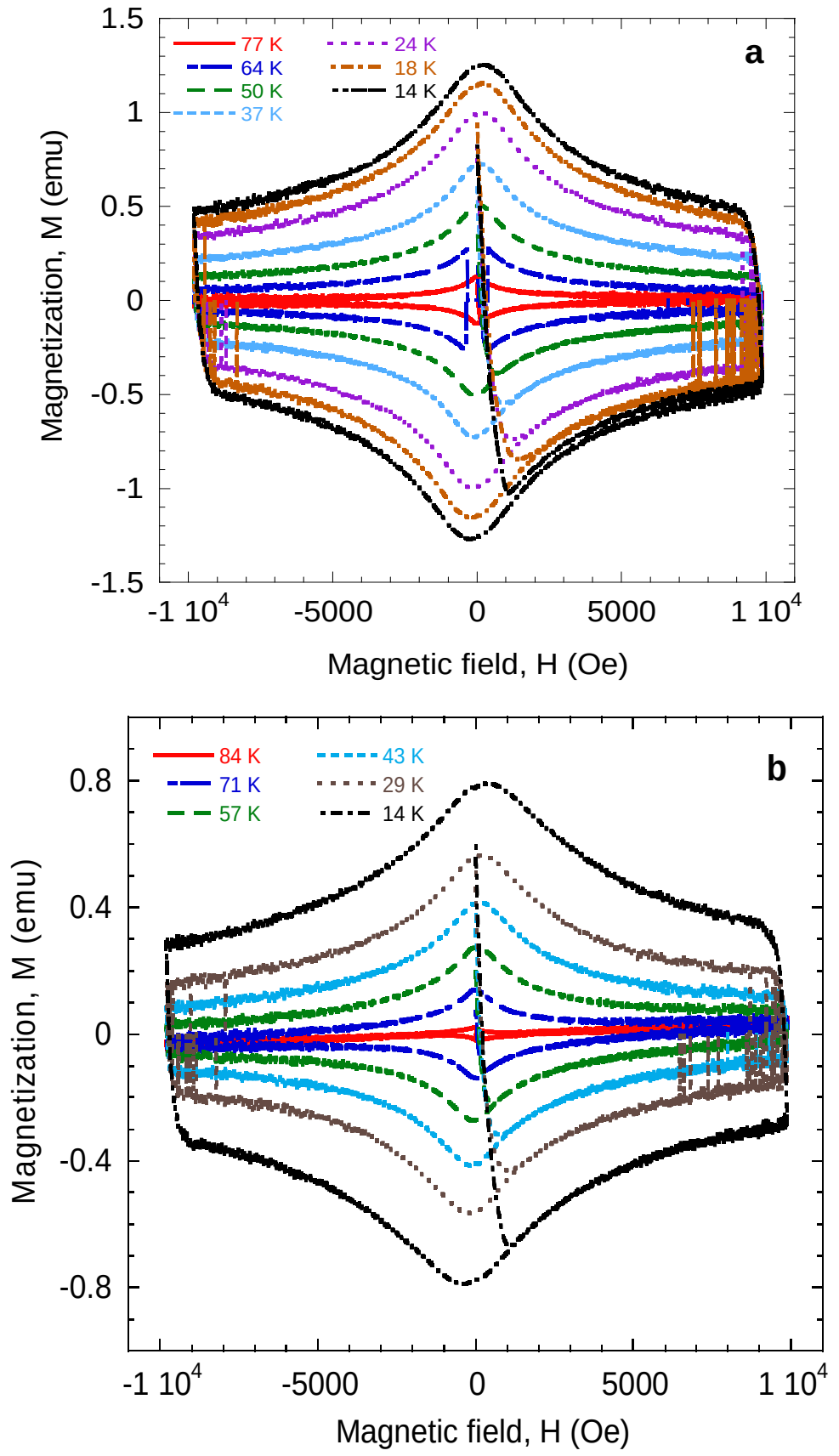

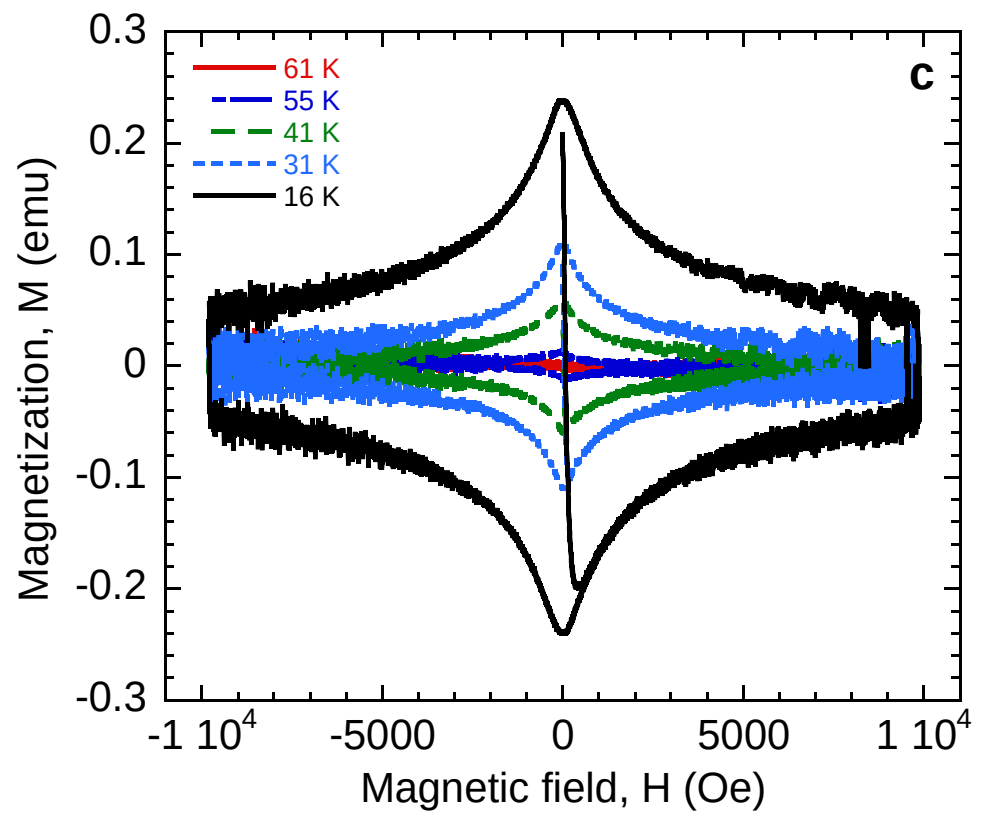

Fig. 3. *M-H* loops for Y123 thin films obtained at different temperatures and hole contents. (a) $p$ = 0.162, (b) $p$ = 0.147 and (c) $p$ = 0.104. For clarity, *M-H* data at a number of other fixed temperatures are not shown in these plots.

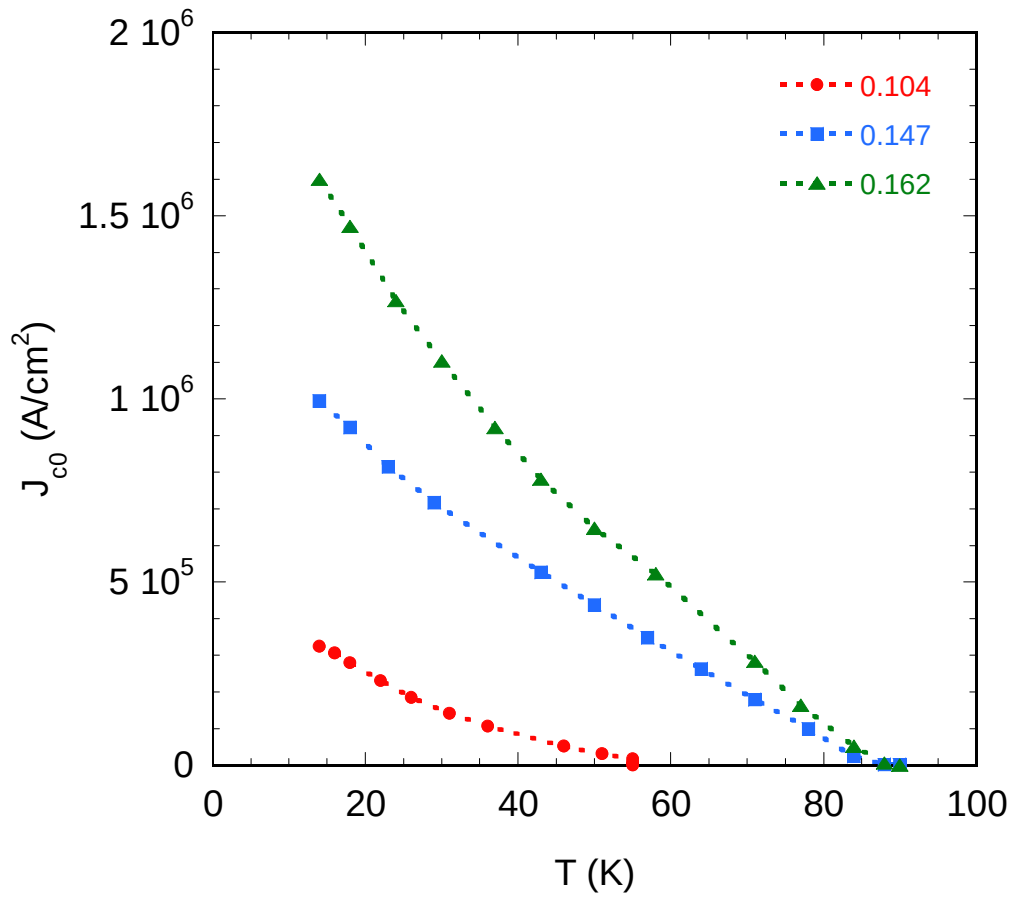

Fig. 4. $J_{c0}(T)$ plots with different values of $p$. The dotted lines are drawn as guides to the eyes.

## 3. Analysis of $J_{c0}(T)$ data

For the analysis, $J_{c0}(T)$ data can be described by a power law temperature dependence, with the power exponent, $n$, depending on the details of the flux pinning mechanism [15]. We have fitted the $J_{c0}(T)$ data using the following equation-

$$J_{c0}(\tau) = J_0(1-\tau)^n \quad (2)$$

where, $\tau = (T/T_c)$, is the reduced temperature and $J_0$ is the extrapolated zero-field critical current density at $T = 0$ K. Results of the fits are shown in Fig. 5.

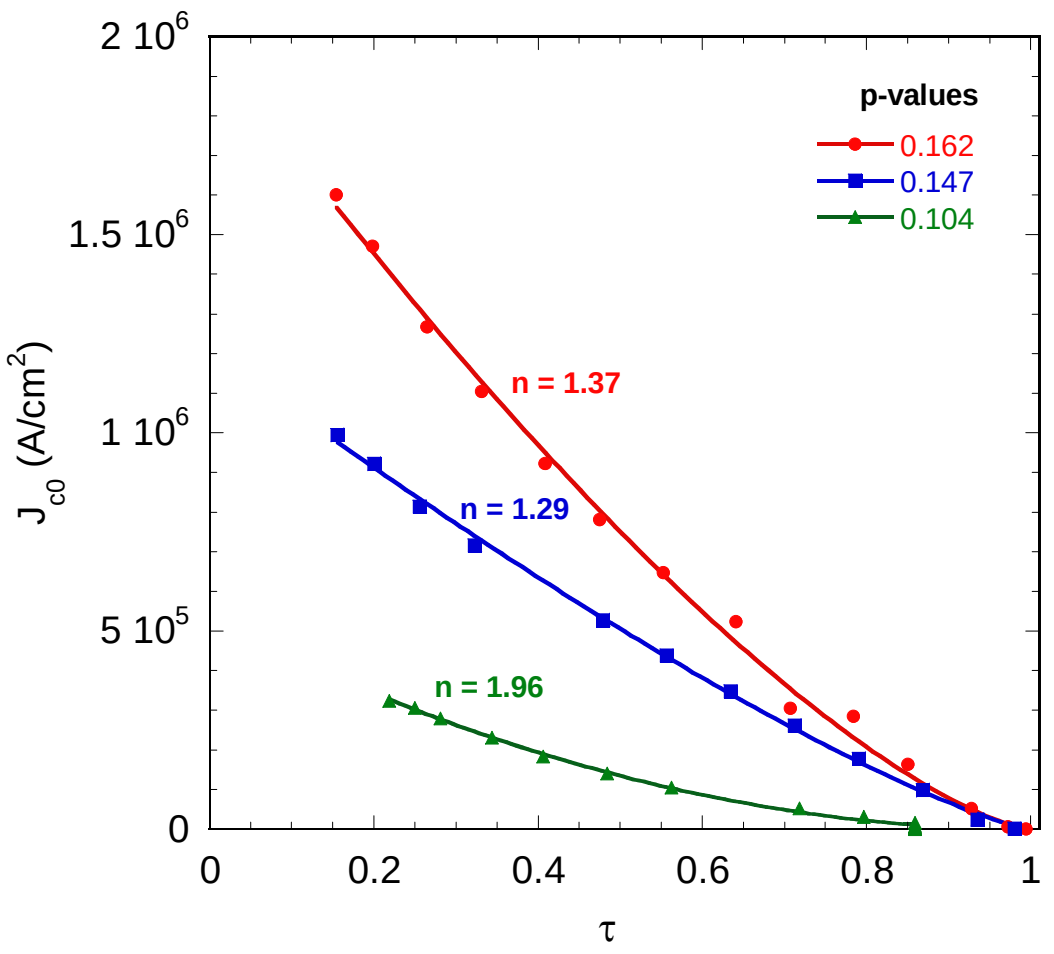

Fig. 5. Fits (full lines) to $J_{c0}(T)$ data. The values for $p$ and $n$ are given in the plot.

The depairing critical density depends on the phase stiffness of the superconducting order parameter and consequently on the superfluid density. Therefore, the temperature dependence of the depairing term should follow the $T$-dependence of $1/\lambda^2(T)$ [16], where $\lambda$ is the (in-plane) London penetration depth. Previous studies [17, 18] found an almost linear temperature dependence of $1/\lambda^2(T)$ for pure Y123 superconductors. It is seen from Fig. 5 that none of the Y123 films under study shows a completely linear $J_{c0}(T)$ even though the ones with small values oxygen deficiencies (0.10 and 0.20) have exponents close to 1.0. The heavily underdoped sample ($\delta$ = 0.31), on the other hand, is characterized by a large exponent ($n \sim 2.0$) and significantly diminished critical current density. These indicate that the observed critical current is governed by both intrinsic and extrinsic effects in all the samples but the relative contributions depending significantly on hole content/oxygen deficiency. To explore the matter further, we have looked for the possible cross-over like behavior in the $T$-dependence of $J_{c0}$. Fig. 6 shows the results of this analysis. From Fig. 6, it is seen that the cross-over features are rather weak for the optimally and the heavily underdoped (HUD) compound. Whereas for the moderately underdoped compound (MUD), there is a clear indication of the change in the power law describing the temperature dependence of the critical current density at a certain temperature (marked by an arrow in Fig. 6). The large value of $n$ together with no clear sign of a cross-over for the HUD ($p$ = 0.104) compound indicates that the critical current density here is primarily governed by the depinning effect. The significantly diminished value of $J_{c0}(T)$ and $J_0$ also suggests that the large amount of oxygen disorder present in this film is not very effective in pinning of the trapped magnetic vortices inside.

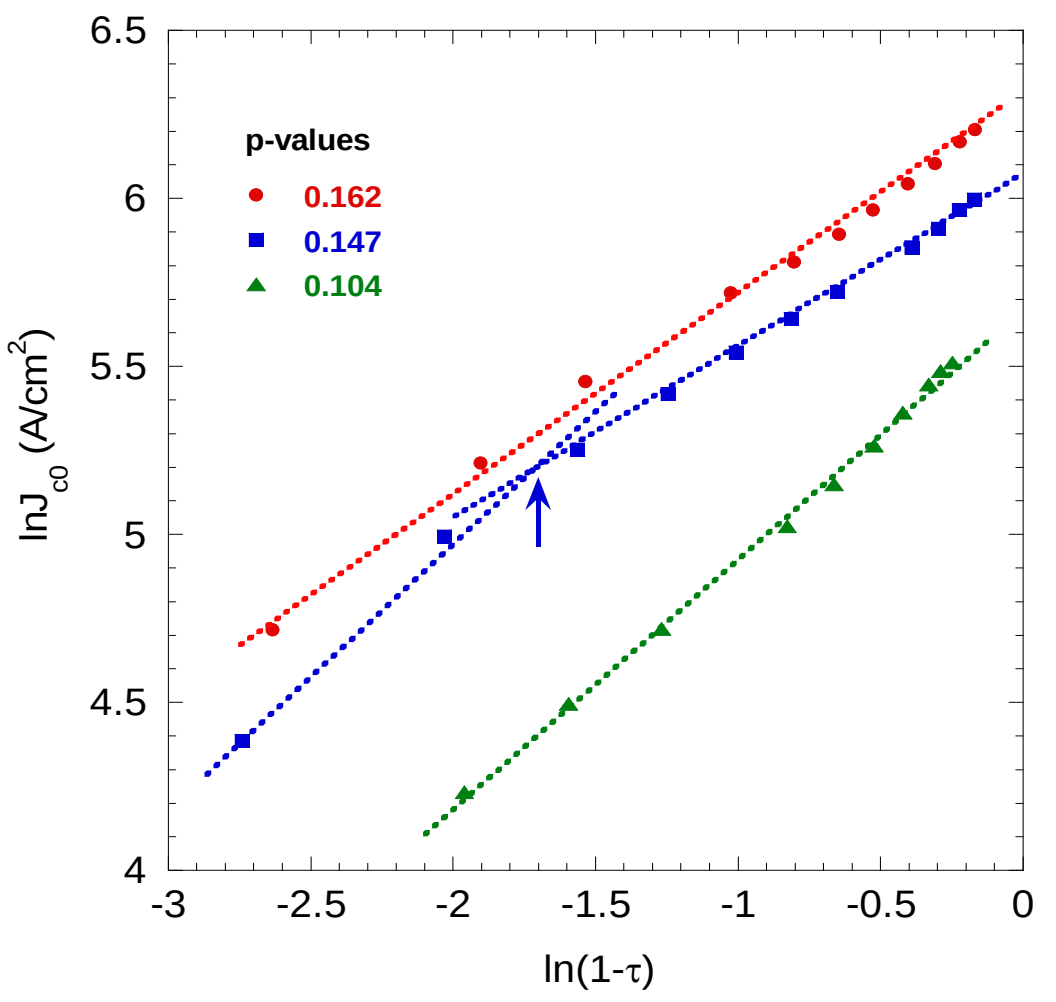

Fig. 6. Possible cross-over in $J_{c0}(\tau)$ data (see text for details). The arrow marks the point where the slope changes significantly and a different flux pinning mechanism comes into play.

It is interesting to note that the high-$T$ (above the cross-over temperature) slope of the In$J_{c0}$ vs. ln(1-$\tau$) plot of the MUD ($p$ = 0.147) compound is almost identical (see Fig. 6) to that for the HUD one. This shows that the depairing critical current dominates at low-$T$ and depinning effect takes over at higher temperatures (characterized by a large slope with $n \sim 2.0$). The probable cause might be the largely reduced phase stiffness at higher $T$ for this MUD compound. For the optimally doped (OPD) film with small oxygen disorder, the critical current density is primarily determined by the depairing effect as no significant sign of a cross-over can be found and $n$ remains small over the experimental temperature range.

## 4. Discussion and conclusions

The extracted values of $J_0$ from the fits to equation 2 are as follows: $1.985\times10^6$ A/cm$^2$, $1.254\times10^6$ A/cm$^2$, and $0.534\times10^6$ A/cm$^2$, for the $p$ = 0.162, 0.147, and 0.102 thin films, respectively. It should be noted that the superconducting transition temperatures for the OPD and the MUD compounds are very similar, 90.5 K and 89.8 K, respectively. $J_0$ for these two films, on the other hand, are very different. $J_0$ of the OPD compound is almost 60% larger than that for the MUD one. The reason for these very different critical current densities in samples with almost identical $T_c$ is related to the very different superfluid densities. The OPD one has a small pseudogap (PG) whereas the PG in the MUD one is significant [19 - 21]. A large PG in the electronic spectral density reduces the superfluid density very effectively. All these findings suggest that (i) a high-$T_c$ itself does not guarantee a high critical current density and (2) oxygen defects in the $CuO_{1-\delta}$ chains are not very efficient in pinning magnetic flux lines. Therefore, to achieve maximum $J_c$ for a given family of cuprates, one must try to increase the superfluid density as much as possible to strengthen the depairing contribution to the critical current. A similar proposal was made by Tallon *et al.* [22] in an earlier study based on collective flux pinning model.

To summarize, we have studied the temperature dependent zero-field critical current density of high-quality *c*-axis oriented Y123 thin films in this paper. Our study reveals that the superfluid density is the prime factor in enhancing the critical current density, a high-$T_c$ alone does not ensure a high $J_c$, oxygen defects act only as weak pinning centers, and depending on the hole content and oxygen deficiency, the critical current density is determined by either depairing or depinning effects. In cases with intermediate oxygen deficiency (*e.g.*, in the MUD film) both these effects may contribute, exhibiting a cross-over behavior at certain temperature.

**Acknowledgements**

SHN thanks the MacDiarmid Institute for Advanced Materials and Nanotechnology, New Zealand, and the IRC in Superconductivity, University of Cambridge, UK for funding this research. SHN also thanks Dr. Anita Semwal for her help with the VSM measurements.